\documentclass[12pt,preprint]{aastex}

\newcommand{\kms}{km\ s$^{-1}$}

\newcommand{\Msun}{$M_{\sun}$}

\newcommand{\Ha}{H$\alpha$\ }
\newcommand{\bm}{beam$^{-1}$}
\newcommand{\yr}{yr$^{-1}$}

\slugcomment{Accepted for Publication in AJ, November 2002}

\shorttitle{Star formation in BIMA SONG Bars}
\shortauthors{Sheth et al.}
\begin{document}

\title{Molecular Gas and Star Formation in Bars of Nearby Spiral Galaxies}

\author{Kartik Sheth\altaffilmark{1}} 
\affil{Division of Mathematical \& Physical Sciences, California
Institute of Technology, Pasadena, CA 91125 \& Department of
Astronomy, University of Maryland, College Park, MD 20742-2421}
\email{kartik@astro.caltech.edu}

\author{Stuart N. Vogel}
\affil{Department of Astronomy, University of Maryland, College Park,
MD 20742-2421} \email{vogel@astro.umd.edu}

\author{Michael W. Regan\altaffilmark{1}}
\affil{Space Telescope Science Institute, Baltimore, MD 21218}
\email{mregan@stsci.edu}

\author{Peter J. Teuben, Andrew I. Harris}
\affil{Department of Astronomy, University of Maryland, College Park,
MD 20742-2421}
\email{teuben@astro.umd.edu}
\email{harris@astro.umd.edu}

\and

\author{Michele D. Thornley\altaffilmark{1}}
\affil{Department of Physics, Bucknell University, Lewisburg, PA 17837}
\email{mthornle@bucknell.edu}

\altaffiltext{1}{Visiting Astronomer, Kitt Peak National Observatory.
KPNO is operated by AURA, Inc.\ under contract to the National Science
Foundation.}

\begin{abstract}

We compare the distribution of molecular gas and star formation
activity in the bar region of six spirals (NGC 2903, NGC 3627, NGC
4321, NGC 5457, NGC 6946, \& IC 342) from the BIMA Survey of Nearby
Galaxies (SONG). The molecular gas, traced using the CO (J=1--0)
emission line, is brightest along the leading edge of the stellar bar
in the bar dust lanes.  The star formation activity, traced using the
\Ha emission line, is offset towards the leading side of the CO
emission.  A cross-correlation analysis shows that a) the HII regions
are offset 0--800 pc on the leading side of the CO emission, b) the
largest offsets are found in the strongest bars, and c) there is a
wide range in offsets in a single bar with no systematic pattern as a
function of the galacto-centric radius.  The CO-\Ha offset constrains
how stars may form depending on the gas flow.  We examine possible
star formation scenarios in context of the two main classes of bar gas
flow simulations, the N-body/sticky particle and hydrodynamic models.
Though both model gas flows are generally consistent with the observed
offsets, we suggest the inclusion of a two- or multi-phase medium to
improve the agreement between models and observation.

\end{abstract}


\section{Introduction} \label{intro}

Barred spiral galaxies are ideal laboratories for the study of star
formation because they host a variety of environments with distinctive
star formation activity and gas dynamics. These environments include
the circumnuclear, inner and outer rings (see discussion of rings in
\citealt{buta96,regan02}), the bar ends, and the bar region itself,
located in between the bar ends and the nucleus. From region to
region, the star formation activity can vary dramatically: bars have
star formation rates of $\sim$0.1--0.4 \Msun \yr\ (e.g.,
\citealt{martin97}), whereas in circumnuclear rings, in an area 10-30
times smaller, star formation rates may be as high as 1 \Msun \yr\
(e.g., \citealt{buta00a}).  Since star formation occurs in molecular
clouds, comparative studies of the distribution of molecular gas and
star formation activity in different environments can shed light on
how star formation may be induced or inhibited.  

Along the spiral arms, for example, observations find that the
molecular gas and HII regions are usually not co-spatial; HII regions
are preferentially offset towards the leading side of the molecular
gas or dust lanes \citep{vogel88,rand93,knapen96,loinard96}.  This
offset is interpreted as evidence of star formation induced by a
spiral density wave \citep{vogel88,rand93}.  However such offsets are
not universally present.  In M100, \citet{sempere97} find that the
offset is absent, or even inverted, along the spiral arms.  Still, in
other cases, the offset is more pronounced between the HII regions and
the dust lanes as the molecular gas and dust lanes diverge (e.g.,
\citealt{lord91,rand99}). \citet{rand99} attribute the divergence to
heating of the gas by young stars or cosmic rays, or to a two-phased
molecular gas medium, but admit that neither of these explanations
works satisfactorily.

In barred spirals, most previous studies comparing the molecular gas
distribution and star formation activity have focused on the highly
active circumnuclear region (e.g.,
\citealt{eckart91,roy93,kenney93,sakamoto95,benedict96}).  In this
paper, we focus on the unique region between the bar ends and the
nucleus. This region is unlike any other in the galactic disk because
it is dominated by highly elliptical stellar orbits (see reviews by
\citealt{sellwood93,a92a,a92b}). Throughout this paper, we refer to it
simply as the {\sl bar}.

Only one previous study has studied the location of HII regions
relative to the stellar bar \citep{martin97}; they found that the
``\Ha bar'' was usually offset towards the leading side of the stellar
bar with misalignments as large as 15$^{\circ}$.  Studies of molecular
gas or dust in bars have found that the gas and dust are also on the
leading side of the stellar bar (e.g.,
\citealt{ondrechen85,handa90,regan95,downes96,sheth00}).  But the
relationship between the molecular gas (or dust) and star formation
has only been studied in a few cases.  In M101, for example,
\citet{kenney91} note that the molecular gas and \Ha emission are at
the same position angle.  In contrast, \citet{sheth00} found that the
HII regions in the bar of NGC 5383 are offset towards the leading side
of the bar dust lanes.  It is unclear whether such offsets are common
in bars.  Since bars have distinctive (and well-studied) gas
kinematics, a study of the relative distribution of the gas and stars
can further elucidate the complex phenomenon of star formation.  With
this goal, we have studied six barred spirals from the recently
completed BIMA Survey of Nearby Galaxies (SONG) \citep{regan01,
helfer02b}.

The sample selection is discussed in \S \ref{sample}, and the
observations and data reduction in \S \ref{obs}.  Using the CO (J=1-0)
emission line to trace the molecular gas, and the ionized hydrogen
(\Ha) line to trace recent star formation activity, we compare the
relative distribution of the two in \S \ref{qual}.  In all six
galaxies we find that the majority of the \Ha emission is offset
towards the leading side of the molecular gas.

We quantify the offset using one and two dimensional cross-correlation
analysis in \S \ref{quant}.  The CO-\Ha offset constrains how stars
may form depending on the gas flow into the dust lane.  We discuss the
results of the cross-correlation analysis in the framework of the two
main classes of bar gas flow models (the N-body sticky particle and
hydrodynamic models) in \S \ref{disc}, and summarize our results in \S
\ref{conc}.

\subsection{Sample Selection} \label{sample}

In BIMA SONG (see details of the survey in
\citealt{regan01,helfer02b}), we detect CO emission in 27 of the 29
barred spirals (e.g., Figures A1 and A2 in
\citealt{sheth01}). Typically, the emission is detected in the
circumnuclear region, where it is usually the brightest. In some bars,
CO emission is also detected at the bar ends, along the bar, and even
in inner rings \citep{regan02}.  Since we are mainly interested in
studying star formation in the bar region we limit ourselves to those
BIMA SONG galaxies in which {\em both} CO and \Ha emission are clearly
detected over a significant portion of the bar. Six galaxies satisfy
this criterion: NGC 2903, NGC 3627, NGC 4321, NGC 5457, NGC 6946 and
IC 342. Global properties of these six are listed in Table \ref{tab1}.

Though small, our sample spans a range of Hubble types with 1 Sb, 2
Sbc and 3 Scd galaxies. This range may be important because most of
the observed differences in star formation activity in bars occur
between early and late Hubble type galaxies (e.g.,
\citealt{elmelm85,ohta86}). For instance, early Hubble type bars have
low star formation activity along the bar and high star formation
activity at the bar ends, whereas late Hubble type galaxies have
higher star formation activity in the bar, but have a gap in star
formation at the bar ends \citep{phillips96}.  

All the galaxies in our sample are classified as SAB in the RC3
catalog. However this does not mean that they are all weak bars
because the Hubble classification of SAB types is not rigorous; in
fact, a recent analysis of infrared data has shown that the true
fraction of strong bars may be as high as 56\%, higher than the
typically quoted 33\% in the RC3 \citep{eskridge00}.  The strength of
a bar may be correlated with its Hubble type.  \citet{elmelm85}
concluded that early Hubble type galaxies have stronger bars because
these bars are longer, relative to their disks, and have flat
profiles; these galaxies also have strong spiral arm patterns.
However, bar strength is a difficult parameter to quantify; other
structural properties such as the bar ellipticity and bulge size are
also important (see discussion in \citealt{buta00b}).

A good indicator of the bar strength is the shape of the bar dust
lanes because it reflects the gas response to the stellar
bar. \citet{a92b} showed that strong bars have relatively straight
dust lanes whereas weaker bars have curved dust lanes. Extending her
analysis to our sample, we infer that NGC 2903 and NGC 3627 are
strongly barred because of their straight dust lanes. NGC 5457 and NGC
4321 with slightly curved dust lanes are weaker bars. IC 342 has an
even more curved dust lanes indicating an even weaker bar. In NGC
6946, the northern dust lane appears to be straight but the southern
dust lane is curved so it is difficult to classify this bar solely on
the shape of the dust lanes.  \citet{regan95} suggest that if there is
bar in NGC 6946, it is rather weak. Using bar dust lanes as a measure
of bar strength, we find that the three latest Hubble types in our
sample are classified as relatively weak bars.  In summary, our sample
contains a broad range of bar strengths, from strong bars in NGC 2903
and NGC 3627, to intermediate strength bars in NGC 4321 and NGC 5457,
to relatively weak bars in NGC 6946 and IC 342.

\section{Observations and Data Reduction}\label{obs}

\subsection{Molecular Gas Data}

All six galaxies were observed in the CO (J=1--0) emission line with
the BIMA (Berkeley-Illinois-Maryland Association) array and the
NRAO\footnote{The National Radio Astronomy Observatory is a facility
of the National Science Foundation, operated under cooperative
agreement by Associated Universities, Inc. telescope at Kitt Peak} 12m
single dish telescope as part of the BIMA SONG key project. The 44
galaxy SONG sample was chosen with the following criteria:
heliocentric velocity, V$_{HEL}$ $<$ 2000 \kms, declination, $\delta
>$ -20$^o$, inclination, $i <$ 70$^o$, and apparent magnitude, B\_T$<$
11.  The typical data cube has a synthesized beam of 6$\arcsec$, a
field of view of 3$\arcmin$.  In a 10 \kms\ channel, the typical noise
is $\sim$58 mJy \bm.  Further details of the data acquisition and
reduction can be found in \citet{regan01} and \citet{helfer02b}.

\subsection{\Ha Data}

We observed NGC~2903, NGC~3627 and NGC~4321 at the 0.9m telescope at
Kitt Peak\footnote{Kitt Peak National Observatory, National Optical
Astronomy Observatories, which is operated by the Association of
Universities for Research in Astronomy, Inc. (AURA) under cooperative
agreement with the National Science Foundation} on the nights of 4--6
April 1999, with the T2KA 2048$\times$2048 CCD camera in f/13.5 direct
imaging mode. In this mode the camera has a 13$\farcm$1 field of view
with 0$\farcs$384 pixels. In the broad band R-filter, we took three
exposures of 180s each, and in the \Ha filter ($\lambda_0$ = 6573 \AA,
$\Delta \lambda$ = 67 \AA) we took three exposures of 100--140s
each. We divided each frame by a normalized flat-field frame and
combined the resulting frames into one single image for each
corresponding filter. Then from these images we subtracted the mean
sky brightness, and removed cosmic rays using standard routines in the
NOAO/IRAF\footnote{IRAF is distributed by the National Optical
Astronomy Observatories, which are operated by the Association of
Universities for Research in Astronomy, Inc., under cooperative
agreement with the National Science Foundation} software package. We
also corrected all images for atmospheric extinction. Finally, we
registered foreground stars in each image with the Hubble Guide Star
Catalog (GSC) and determined astrometric solutions for each image. The
residuals in determining absolute positions were smaller than
0$\farcs$2; however, systematic errors in the GSC prevent us from
achieving an accuracy $<$ 1$\arcsec$. We used the R-band image to
subtract the underlying continuum from the \Ha image.

We observed IC~342 at the 1.5m telescope at
Palomar\footnote{Observations were made on the 60 inch telescope at
Palomar Mountain, which is jointly operated by the California
Institute of Technology and the Carnegie Institution of Washington.}
on the night of 19 November 1999 with the Maryland-Caltech Fabry-Perot
camera. These data were reduced and calibrated using the procedure
described in \citet{vogel95}. The \Ha image for NGC 6946 is also a
similar velocity-integrated \Ha Fabry-Perot map obtained previously at
Palomar. A broadband continuum image and a continuum-subtracted \Ha
image for NGC 5457 were supplied to us by R. Kennicutt; these images
have 2$\farcs$6 pixels compared to 0$\farcs$38 pixels for the other
optical images.

\section{Results}\label{cohares}

\subsection{Comparing the Distribution of Molecular Gas and Star Formation}\label{qual}

We first consider the relative distribution of the molecular gas and
star formation activity in the six galaxies by overlaying the CO
emission on the continuum-subtracted \Ha images (Figures
\ref{2903olay}--\ref{ic342olay}). Our goal is to learn where the stars
may be forming by comparing the relative locations of molecular clouds
and HII regions.  Since any image is only a snapshot in time it is
difficult to assign a star forming region to its parent molecular
complex.  We make the usual assumption that CO traces regions most
prone to star formation, and \Ha traces young stars most likely to
have formed in molecular gas complexes traced by the CO.

In every image we show the extent of the stellar bar, as determined by
\citet{sheth01}, with an ellipse.  The ellipse corresponds to the
outermost bar isophote and was determined from fitting ellipses to
optical or near-infrared images following a method devised by
\citet{reganelm97}.  The width of a bar is a difficult parameter to
measure accurately because contamination by bulge light tends to make
the fitted ellipses wider \citep{sheth01}.  Since the minor axis
length is not relevant for the study presented here, we chose a
representative minor axis that is one-quarter that of the major axis
for all bars except IC 342, which has a small, oval bar for which we
chose a minor axis that is one-half that of the major axis.

In each image (Figures \ref{2903olay}--\ref{ic342olay}) we show CO
contours overlaid on a grey-scale image of the continuum-subtracted
\Ha emission.  We describe the distribution of \Ha and CO using
approximate position angles because their morphology is often curved.
Compact \Ha structures are labeled with a prefix H, and CO features
are labeled with a prefix C.  We also show the direction of rotation
of the bar with a curved arrow symbol in the lower right hand corner;
the direction is based on the assumption that spiral arms trail in
galaxies. All the galaxies in our sample rotate counter-clockwise
except for NGC 6946.

In all six galaxies we find that the CO emission is brightest on the
leading side of the stellar bar.  Overlays of the CO maps on
optical-infrared color maps (not shown) confirm that the molecular gas
generally coincides with the bar dust lanes \citep{sheth98}.  In five
out of the six galaxies, weak spurs of CO emission are seen on the
trailing (upstream) side of the dust lanes.  At the bar ends, bright
CO emission is present on the leading and trailing sides; this gas is
most likely associated with inner rings, which often encircle bars
\citep{buta96,regan02}.  The \Ha emission is distributed in bright,
compact structures and weak, diffuse emission.  The differences in bar
star formation activity noted by \citet{phillips96} are not always
present.  For instance, both the early Hubble type bar in NGC 3627 and
the late Hubble type bar in NGC 5457 have relatively little star
formation activity, whereas strong activity can be seen in the bar of
NGC 2903.  But such differences across Hubble type are difficult to
quantify with this small sample.  In these six bars, while there are
a few cases where \Ha is coincident with, or on the trailing side
of CO, {\sl all six} show that majority of the \Ha emission is
offset towards the leading side of the CO emission.  We describe these
results for each individual galaxy in detail below:

\noindent{\em NGC 2903 (Figure \ref{2903olay})}: The stellar bar in
NGC 2903 is oriented at a position angle (PA) of 24$^{\circ}$ east of
north. The molecular gas is distributed towards the leading side of
the stellar bar, with a PA of $\sim$30$^{\circ}$. The \Ha emission is
offset even farther towards the leading side of the CO emission.

In the northern half of the bar, the CO emission is brightest in a
large complex, labeled C1. A bright \Ha complex, H1, coincides with
part of the C1 complex. 3--4$\arcsec$ south of C1 is a narrow ridge of
CO, labeled C2, which bridges C1 with the bright circumnuclear
gas. Several discrete \Ha features (H2, H3, H4, and H5) are seen on
the leading side of the CO.  These structures are connected by a
diffuse, concave ridge of \Ha emission. On the trailing (upstream)
side, some diffuse \Ha emission is present but there is a conspicuous
lack of HII regions; this is not an artifact of dust extinction.  In
the dust lanes, the dust extinction is usually insufficient to hide
HII regions (e.g., \citealt{regan95}). Also, the gas density is
expected to be dramatically lower on the trailing side of the dust
lane as the gas piles up in the bar dust lane (e.g., see model density
distributions in \citealt{combes85,a92b,piner95}).

In the southern half of the bar, there are two bright CO complexes, C3
and C4. As in the northern half, \Ha complexes (H6, H7, and H8) are
all on the leading side of the CO emission. A diffuse ridge of \Ha
emission extends south of H7, with a position angle of $\sim$5$^{\circ}$,
on the leading edge of the CO emission. In this half of the bar, we
also see a weak CO spur, labeled C5, east of C4 on the trailing side
of the dust lane. Diffuse \Ha emission is present at the trailing edge
of C5. In two places, near H7, and north of H6, there is weak CO
emission on the leading side of the dust lane.

\noindent{\em NGC 3627 (Figure \ref{3627olay})}: The bar in NGC 3627
is oriented at a PA of 161$^{\circ}$.  The molecular gas
is mostly on the leading side the stellar bar, with a PA of
$\sim$164 $^{\circ}$, but note that at the southern end of the bar,
the molecular gas morphology turns sharply towards the trailing side
of the bar. The \Ha emission is offset towards the leading side of the
CO.  The CO intensity decreases from the inner ends of the bar towards
the outer ends, i.e. from C3$\rightarrow$C2$\rightarrow$C1 in the
northern half, and from C4$\rightarrow$C5$\rightarrow$C6 in the
southern half. This behavior is consistent with the predictions of the
hydrodynamic models \citep{sheth00}.

In the northern half of the bar, there is only one bright, compact \Ha
structure, labeled H1, located 30$\arcsec$ north of the nucleus, and
3--5$\arcsec$ east, and towards the leading side of the CO ridge
connecting C1 and C2.  Diffuse \Ha emission connecting H1 to the bar
end is also present on the leading edge of the CO.  As in NGC 2903,
there is a paucity of \Ha emission on the trailing side of the CO.
Additional CO and \Ha emission associated with an inner ring (shown by
the dashed segments) circumscribing the bar, is also present; this
inner ring is discussed in detail in \citep{regan02}.

In the southern half of the bar, all three compact \Ha structures (H2,
H3, and H4) are also on the leading edge of the CO emission,
3--4$\arcsec$ west of the CO ridge connecting C4, C5, and C6. As in
the northern half, diffuse \Ha emission connects H2, H3, and H4; most
of this emission is also on the leading edge of the CO.

\noindent{\em NGC 4321 (Figure \ref{4321olay})}: The bar in NGC 4321
is oriented at a PA of 102$^{\circ}$.  The molecular gas is mostly on
the leading side of the stellar bar with a PA$\sim$90 $^{\circ}$; this
is more clearly seen in the western half of the bar where the CO
emission runs continuously down the entire half of the bar. In the
western half, the \Ha emission is offset another 3-4$\arcsec$ towards
the leading side, with approximately the same PA.  In the eastern half,
the \Ha emission is weak and no clear offset is seen.

In the western half of the bar, a relatively narrow lane (10$\arcsec$
in width) of CO emission emerges from the bright, central
40$\arcsec\times$ 20$\arcsec$ oval-shaped distribution of CO emission.
This lane extends to the west (C1$\rightarrow$C2$\rightarrow$C3) and
then curves inwards to the spiral arms from C3 to C4. As in NGC 3627,
the CO intensity decreases away from the central region.  A spur-like
feature, labeled C5, is seen 5$\arcsec$ south of C1 extending almost
10$\arcsec$ ($\sim$780 pc) on the trailing side of the bar. These
spurs may play an important role in forming stars because they have
high gas density but are located in an area with lower shear
\citep{sheth00}.

The discrete \Ha structures (H1, H2, and H3) are evenly-spaced
($\sim$10$\arcsec$), and are on the leading edge of the CO emission. A
narrow ridge of diffuse \Ha emission connecting these structures is
offset north from the CO by about 4$\arcsec$.  Farther west, towards
the bar end, a bright concentration of molecular gas, labeled C4, is
coincident with the brightest \Ha emission in this half of the bar,
labeled H4.

In the eastern half of the bar, the two main concentrations of CO, C6
and C7, are on the leading side of the stellar bar. Two HII regions, H5
and H6, are offset $\sim$5$\arcsec$ west of C7 and C6
respectively. Unlike the western half, there is no clear leading
offset of HII regions from the CO emission. At the eastern bar end,
bright CO and \Ha emission are present in the regions labeled C8 and
H7.

CO and \Ha emission are distributed on the trailing {\em and} leading
sides at the bar ends (10$\arcsec$ north of C4, and 12-15$\arcsec$
south of C8 and H7). Such a morphology seems to be common in bars and
is most likely associated with inner rings \citep{buta96} that
encircle bars.

\noindent{\em NGC 5457 (Figure \ref{5457olay})}: The bar in NGC 5457
is oriented at a PA of 80$^{\circ}$.  The molecular gas is primarily
on the leading side of the stellar bar with a PA$\sim$95~$^{\circ}$.
The \Ha emission tends to be towards the leading side of the CO
emission.

In both halves of the bar the CO intensity decreases outwards from the
central concentration of gas, to the east towards C1, and to the west
towards C6. In the western half of the bar, there is one bright CO
complex, labeled C2 in Figure \ref{5457olay}.  Generally coincident
with but offset slightly towards the leading side of C2 is a bright
HII region, labeled H1. Diffuse \Ha emission connecting the nuclear
region and H1 is seen mostly on the leading side of the CO emission in
the region in between C1 and C2. To the south of C2, on the trailing
side, there is a 10$\arcsec$ spur-like CO feature, labeled C3.

In the eastern half, the CO morphology is very similar to that seen in
the western half. C7 is a weak CO peak, $\sim$30$\arcsec$ east of the
circumnuclear region. Associated with C7 is weak \Ha emission, labeled
H3. Generally the \Ha emission in this half of the bar is
characterized by a continuous ridge extending from the circumnuclear
emission to H4, on the leading side of the CO. H4, like H2, is on
the leading side of the bar and is probably the beginning of an inner
ring.

Towards both bar ends, the CO emission deviates away from the leading
side of the bar and connects with the CO emission in the spiral arms
of NGC 5457. The CO region labeled C4, and the \Ha region labeled H2,
are probably associated with the bar end, but the two bright CO and
\Ha peaks, 6--10 arcseconds north of H2, are associated with the
spiral arm.

\noindent{\em NGC 6946 (Figure \ref{6946olay})}: The stellar bar in
NGC 6946 is oriented at a PA of 19$^{\circ}$.  In the bar, the
molecular gas is distributed asymmetrically about the center. To the
north of the circumnuclear region, the CO emission covers a rather
broad 40$\arcsec \times$40$\arcsec$ area, with a sharp ridge of CO
emission on the leading side of the bar with a PA of
$\sim$0$^{\circ}$. In the southern half of the bar, the broadly
distributed CO emission is truncated to the south, extending only
20$\arcsec$ south of the circumnuclear area. While the CO and \Ha
emission overlap considerably, the brightest \Ha emission tends to be
on the leading edge of the CO.

In the northern half of the bar, a ridge of CO emission extending from
the circumnuclear region towards C1 gradually decreases in intensity.
Along this ridge there is an HII region, 10$\arcsec$ north of the
nucleus labeled H1, a CO peak, 10$\arcsec$ farther north labeled C1,
and two \Ha peaks, labeled H2 and H3 even farther north of C1.  We
find a significant amount of CO emission on the trailing side of this
bar.  There is a concentration of CO emission, labeled C2, curving
east from C1. An HII region is coincident with it.  On the trailing
side of the bar, there is CO emission C3, C4, and C5, and \Ha peaks at
H4 and H5.

In the southern half a ridge of CO emission extends from the nuclear
region towards C7. Associated with C7 is the brightest \Ha emission in
this half of the bar. Farther south of C7, the CO emission narrows
into a ridge, labeled C9, and then ends in a broader peak at C10. On
the trailing side of C9 is an HII region labeled H8. Two HII regions,
H9 and H10, are seen in the leading side of the bar, a few arcseconds
northeast of C10. There is also an trailing extension of CO emission
to the east of H7, labeled C8.

\noindent{\em IC 342 (See Figure \ref{ic342olay})}: IC 342 has a
small, oval stellar bar.  The bar ellipse shown in Figure
\ref{ic342olay} is at a PA of 28$^{\circ}$.  The CO emission appears
to be in the center of the bar with a PA of $\sim$0--10$^{\circ}$.
This central distribution of gas is unusual because it is expected
only when the bar is extremely strong and inner Lindblad resonances
are absent \citep{a92b}.  Such a situation is extremely short-lived in
the models.  It is more likely that the bar parameters for this galaxy
are misidentified because the bar resides in a bright starburst
circumnuclear region.  A recent study by \citep{cros01} suggests that
although the bar is most prominent in the central 2$\arcmin$, CO
kinematics indicate a gaseous bar as long as 4$\farcm$7 with a
PA$\sim$0$^{\circ}$.  In that case, the CO emission would be on the
leading edge of the bar.

As in other bars, the CO emission along both dust lanes decreases in
intensity outwards from the circumnuclear region towards C2 to the
north, and C4 to the south. In the northern half of the bar there is
an extension of CO, labeled C1, towards the trailing side. A few
arcseconds north of it is an HII region, labeled H1, which also lies
along the leading edge of the CO emission.  Farther north near the
bar end, there is a CO emission peak, labeled C3. A few arcseconds
north and east of C3 is another \Ha peak, labeled H2. 

In the southern half of the bar there is a concentration of CO
emission, labeled C5, on the trailing side of the main lane of CO
emission. There is also a ridge of CO emission which extends southeast
of the nucleus toward the bar end, ending in a CO peak, labeled C6. In
the bar there is a lot of diffuse \Ha emission but the few HII regions
(H3 and H4) are on the leading side of the CO ridge.

In summary, CO emission in all six bars is brightest on the
leading side of the stellar bar in the bar dust lanes. The CO
intensity generally decreases towards the bar ends.  Weak spurs of CO
emission are sometimes seen on the trailing side of the bar.  At the
bar ends, there is often bright CO and \Ha emission on both the
trailing and leading sides; this emission is probably associated with
inner rings.  In a few instances \Ha emission is coincident with, or
on the trailing side of the CO emission.  But in all six bars, the
majority of the \Ha emission is on the leading side of the molecular
gas.

\subsection{Quantifying the Offsets}\label{quant}

The typical distance between molecular gas complexes and HII regions
sheds light on when, where and how stars form. To accurately measure
this distance, we need to trace the orbit of an HII region and measure
the separation along this path. In a typical galaxy disk, this path is
usually circular and therefore a measurement of the azimuthal
displacement is sufficient. However non-axisymmetric
perturbations such as bars and spiral arms can introduce streaming
motions in the gas (e.g., \citealt{roberts79,a92b} and references
therein). In these cases, an orbit has both an azimuthal and a radial
component. The radial displacement depends on the strength of the bar
or spiral arm, and it may vary with radius. We expect that the radial
streaming motions will be most important at the inner ends of the bar
and spiral arms \citep{roberts79}.  From \S \ref{qual}, we infer that,
to first order, the displacement is mostly in the azimuthal
direction. Therefore we first quantify the distance between the CO and
\Ha by doing a one-dimensional cross-correlation in azimuth (as
described in the next paragraph), and then we perform a
two-dimensional (radial + azimuthal) cross-correlation over the entire
bar region.

\subsubsection{Cross-Correlation Analysis}\label{oned}

We re-gridded each \Ha image to the CO image using standard
MIRIAD routines \citep{sault95}, deprojected both images using ZODIAC
\citep{miyashiro82, shopbell97} routines developed by ourselves and
\citet{gruendl96}, and then mapped the images on to a polar grid. The
deprojected CO (top left) and \Ha (bottom left) images are shown in
the left column of Figures~\ref{n2903all}--\ref{ic342all}.  For the
one-dimensional cross-correlation analysis, we cross-correlated
1$\arcsec$ annuli. The result of these analyses are shown in the
vertical panel on the right hand side of
Figures~\ref{n2903all}--\ref{ic342all}. The star symbols indicate the
angular offset corresponding to the maximum of the cross-correlation
for that radius. Note that this value averages the offset between the
CO and \Ha in the two halves of the bar. The solid circles on the
deprojected \Ha and CO images are the boundaries within which the
cross-correlation is done. The inner radius is chosen to best avoid
the circumnuclear emission and the outer radius is at the end of the
bar (corresponding to the bar ellipse in Figure
\ref{2903olay}--\ref{ic342olay}). The dashed circles are typically
spaced at integral multiples of 10$\arcsec$ and the corresponding
radius is shown with a solid, horizontal line in the one dimensional
cross-correlation diagrams in the right hand panel.

The cross-correlation result can be affected by systematic dust
extinction. For example, consider a barred spiral where the \Ha
emission is distributed evenly on the trailing and leading sides. The
cross-correlation peak of such a distribution should not show any
azimuthal offset. However, if the \Ha on the trailing side of the dust
lane is obscured by a screen of dust, then the cross-correlation
diagram would show that the \Ha is preferentially on the leading side
of the dust lane.  But as noted earlier in \ref{qual} this is unlikely
because the dust extinction on the trailing side is very low as gas
piles up in the dust lanes.  Note that any extinction suffered by the
\Ha in the dust lane only increases the observed offsets; so the
CO-\Ha offsets calculated here are upper limits.

It is important to understand whether the cross-correlation peak
depends on small, bright structures, like the \Ha and CO peaks, or on
large, diffuse, and dim features.  We performed two tests to address
this question.  First, we flagged all low level emission in the CO and
\Ha maps of NGC 2903 and cross-correlated the images in radius and
azimuth. We found that the cross-correlation peak was essentially
unchanged. The main difference was that a background of small but
non-zero correlation values, present in the original cross-correlation
diagram, was now removed.  The diffuse emission, therefore, is
analogous to a constant offset contributing to the correlation
diagram, but at no particular lag.  This test indicates that the
cross-correlation peak is determined primarily by the bright peaks in
the maps.

As a second test, we set all emission above an arbitrary bright signal
value to that signal value in the \Ha and CO maps for NGC 2903,
approximating the removal of bright peaks. In the resulting
correlation diagram there was a trend of \Ha offset azimuthally
towards the leading side from the CO but the cross-correlation peak
moved significantly from its original location and became broader.
The larger width of the peak was expected because we removed the
sharpness inherent in the maps being correlated. The shift, once
again, showed that the location of the correlation peak is primarily
determined by bright \Ha and CO peaks.  Assuming that these peaks
trace regions most prone to star formation, and recently formed OB
associations respectively, the correlation peak effectively quantifies
the offset between parent molecular gas complexes and newly formed
stars.

\subsubsection{Results of the One-Dimensional Cross-Correlation}\label{onedres}

In every bar (right hand panels in Figures
\ref{n2903all}--\ref{ic342all}), we find that the peak of the
cross-correlation shows \Ha azimuthally offset towards the leading
side of the CO emission, consistent with the discussion in \S
\ref{qual}. In these panels, we have annotated the correlation peaks
with the notation used in Figures~\ref{2903olay}--\ref{ic342olay} for
the corresponding CO and HII regions.

The offset between the annotated regions varies from 0--800 pc with
a range of offsets in any given bar.  We find that the largest offsets
are generally seen in the strongest bars.  For instance, in NGC 2903,
offsets of $\sim$800 pc are seen in the northern half between C2 \&
H5, and between H4, H5 \& the northern part of C1.  In the same bar,
smaller offsets are also present between C3 \& H6 ($\sim$400 pc), and
between C1 \& H1 ($\sim$200 pc).  In the slightly weaker bars, NGC
4321 and NGC 5457, the HII regions are typically offset 200-500 pc
(e.g., C1 \& H1 in Figure \ref{n4321all}), and in the weakest bars,
NGC 6946 and IC342, the offsets are even smaller, typically less than
400 pc.

At some radii, the cross-correlation peak is at larger distances,
exceeding a kiloparsec in some cases. These are instances where only
weak, and diffuse \Ha emission is being correlated with the CO
emission.  An example of this is seen between 20--25$\arcsec$ in NGC
3627 (Figure \ref{n3627all}).  In a few cases we get non-physical
values for the cross-correlation peaks; this occurs when we correlate
either noise features or weak emission that is well outside the bar.
The arrows point towards the non-physical value of the
cross-correlation peak, off the scale shown in these figures.

Though there appears to be a relationship between the bar strength and
the CO-\Ha offset, there is no {\em systematic} pattern in the offset
as a function of the galacto-centric radius.  We often see a trend of
increasing offsets of \Ha from CO over radial extents of
5--10$\arcsec$, e.g., the radial bins between 38--42$\arcsec$ (C2 \&
H5) in Figure \ref{n2903all}.  This trend is due to beam smearing and
subsequent deprojection of the images.  These results are discussed
further in \S \ref{disc}.

\subsubsection{Results of the Two-Dimensional Cross-Correlation}\label{twod}

The two-dimensional correlation results are plotted in the bottom
panel in Figures~\ref{n2903all}--\ref{ic342all}. In these panels, the
intersecting straight lines indicate the 0,0 lag position. Positive
(negative) correlation values indicate \Ha offset radially inwards
(outwards) or azimuthally leading (trailing) the CO. Unlike in the
one-dimensional analyses, where we only correlated 1$\arcsec$ annuli,
the location of the cross-correlation maximum (or maxima) in the
two-dimensional analyses is susceptible to bright emission near the
circumnuclear or bar-end regions because we sum over the whole bar
region. While we did our best to choose an inner boundary to avoid
most of the circumnuclear emission, the outer boundary sometimes
included bar-end/spiral arm emission (for instance, in NGC 2903).
This is one of the reasons why the two dimensional cross-correlation
results are less reliable than the other analyses presented here.
Another is the possibility of getting incorrect results because we may
correlate physically unrelated peaks.  Suppose the brightest HII
region is at the outer bar end whereas the brightest CO peak is at the
inner bar end.  Then the cross-correlation peak will be at the radial
lag at which these peaks overlap.  But these regions are most likely
unrelated.  Though we restricted the radial lag space to avoid such
extreme situations, one must treat the two dimensional
cross-correlation results with caution.

In all six galaxies, the \Ha is leading the CO emission in the
azimuthal direction.  In other words, the correlation values are
higher on the right hand side of the bottom panel.  This is consistent
with the overlays (\S \ref{qual}) and the one-dimensional
cross-correlation analyses (\S \ref{oned}).  We also find that, in
almost all cases, there are higher cross-correlation values for \Ha
offset radially outwards from the CO.  In four galaxies (NGC 2903, NGC
3627, NGC 6946 and IC 342), the maximum of the cross-correlation
values is located at outward radial lags of $\sim$7-10$\arcsec$. These
large offsets are not obvious in the overlays of CO and \Ha images,
and may not be physically meaningful because, as discussed above, the
correlation is between \Ha and CO peaks along the dust lane.  

\section {Discussion}\label{disc}

How can one explain the CO-\Ha offset?  If there was no relative
motion of the gas and stars, then the CO and \Ha should be co-spatial.
However, since the majority of the \Ha emission is preferentially on
the leading side of CO in all six bars, gas dynamics must play a role.
Hence, we consider the two main classes of bar gas flow models: a)
N-body/sticky particle models which treat the gas as a collection of
sticky particles, reacting to a gravitational field, and exchanging
angular momentum and energy upon collisions (e.g.,
\citealt{combes85}), and b) grid-based hydrodynamic models which treat
the gas as an isothermal, ideal fluid obeying standard hydrodynamic
equations (e.g., \citealt{a92a,a92b,piner95}).  We use the framework
provided by these models to understand the observed CO-\Ha offsets.

\subsection{An Important Caveat: Detection Limits on the Molecular Gas Data}\label{det}

The addition of single dish (NRAO 12m) On-The-Fly data to the BIMA
interferometric observations greatly reduced the problem of spatial
filtering.  As discussed by \citet{helfer02a} and \citet{helfer02b},
the combined data, such as those presented here, recovers 80-90\% of
the flux density for all sources of size $<$ 10$\arcsec$, with the
recovery improving for smaller scales. At the mean distance to these
six galaxies (8.25 Mpc), 10$\arcsec$ subtends a linear scale of 400
pc, much larger than a typical giant molecular cloud ($\sim$40--50 pc,
\citealt{scoville90}), and larger than most giant molecular complexes
or associations ($\sim$100-400 pc, \citealt{rand93}).  So these data
do not suffer greatly from the spatial filtering of an interferometer.

The sensitivity of these observations is discussed in detail in
\citet{helfer02b}.  We summarize the main points here.  Given our
typical rms noise of 58 mJy \bm, we have a rms column density
sensitivity of 4.6 \Msun\ pc$^{-2}$, using a CO/H$_2$ conversion
factor of 2$\times$10$^{20}$ cm$^{-2}$ (K \kms)$^{-1}$ in a 10 \kms\
channel.  The data quality is such that the edges of extended emission
are probably reliable at 2$\sigma$ or 9.1 \Msun\ pc$^{-2}$, and
isolated points are reliable at 3$\sigma$ or 13.7 \Msun\ pc$^{-2}$.
For comparison, the mean surface density of a Galactic giant molecular
cloud in the solar neighborhood is 50--100 \Msun\ pc$^{-2}$
\citep{blitz93}, within an area of 2.1 $\times$ 10$^3$ pc$^2$.
Assuming that this surface density is typical for molecular gas, we
are sensitive to clouds with typical masses $\sim$ 10$^6$ \Msun\ in a
single beam, or in other words, we would detect large molecular clouds
in a single beam but our observations would not detect a 10$^5$ \Msun\
cloud.  Hence it is possible that we are not detecting molecular gas
from small, weak clouds that may be associated with the observed HII
regions.  This caveat applies throughout our discussion.

Our analysis is focussed on the observed offset between the CO and \Ha
emission.  The basic assumption is that these observations trace the
highest surface density CO emission, i.e., the regions most prone to
star formation activity.  In the next two sections, we discuss how
star formation may occur given the gas flow in the N-body/sticky
particle and hydrodynamic models.

\subsection{Gas flow and Star Formation in the N-body/sticky Particle Models}

The N-body/sticky particle simulations (e.g., \citealt{combes85}) are
designed to emulate the behavior of bound, molecular clouds, the basic
organizational unit of the molecular interstellar medium.
However, these models parameterize the physics involved by using
ad~hoc equations for simulating the collisions and energy
exchange. Though they have been successful at reproducing curved dust
lanes and rings \citep{combes96}, they are unable to reproduce the
sharp velocity jumps and straight dust lanes observed in strongly
barred spirals.

In these models, the bar dust lane results from orbit crowding,
similar to a spiral arm dust lane (see Figure 10 in
\citealt{regan99}). Gas clouds in these models crowd together and
probably form giant molecular complexes in the dust lane; they
eventually leave the dust lane diverging outwards on the leading side
of the bar. If stars form in the molecular complexes, as suggested for
the spiral arm of M51 \citep{vogel88}, then we naturally expect \Ha
emission to be both in the dust lane and downstream from it. Thus,
these models naturally fit the observed CO-\Ha offset seen in our
sample. They may also explain the presence of \Ha emission on the
trailing side of the dust lane by arguing that, as the clouds
gradually converge on the trailing side they may collide, form
complexes, and form stars.  Also as noted by \citep{vogel88}, the
molecular gas clouds are difficult to detect on the leading
(downstream) side of the dust lane because the average gas surface
density decreases as the clouds diverge outwards from the dust lane.
These models cannot easily explain the radially outwards \Ha offset
seen in the two dimensional cross-correlation diagrams because, over
most of the bar, the cloud orbits have a radially inwards component as
clouds converge towards the dust lane.

\subsection{Gas flow and Star Formation in Hydrodynamic Models}

In contrast to the N-body/sticky particle models, hydrodynamic
simulations use standard fluid equations to model the gas but the gas
is usually an isothermal, ideal fluid.  Though self-gravity and
magnetic fields have not been included in the past, newer models have
begun to incorporate these as well (e.g., \citealt{kim02}).  In
contrast to the N-body/sticky particle models, the hydrodynamic models
have been successful, not only at reproducing the observed gas
kinematics \citep{regan97,regan99}, but also the entire range of
observed bar dust lane shapes, from curved to straight \citep{a92b}.

The gas kinematics in these models are dramatically different (see
Figure 9 in \citealt{regan99}). The gas streamlines diverge as they
approach the dust lane \citep{regan97}. At the dust lane, {\sl all} of
the gas is redirected inwards by a hydrodynamic shock such that {\em
none} of the gas crosses the dust lane.  Historically, the bar dust
lanes, despite their high gas density, are regarded as inhospitable
environments for star formation because of high shear \citep[and
references therein]{a92b}.  \citet{regan97} have also argued that the
diverging streamlines on the trailing side of the dust lane can tear
apart molecular clouds, or at least prevent molecular complexes from
forming (but note that self-gravity can counteract this effect, see
discussion in \citealt{sheth00}).  These studies argue effectively for
reduced star formation efficiency in the dust lanes.

Even if stars form in the dust lane, the gas flow dictates that
stellar clusters travel down the dust lane.  Thus these models cannot
naturally explain the azimuthal offset between the \Ha and the CO.  In
this model gas flow, the only way for the HII regions to appear on the
leading side of the dust lanes is to form stars on the trailing side
while the gas still has tangential motion.

Such a mechanism was proposed in a previous study of the barred spiral
NGC 5383.  In that bar, \citet{sheth00} noted that the limited number
of HII regions, all on the leading side of the dust lane, were
directly across from faint dust spurs.  They proposed that, in this
bar, stars were forming in spurs because spurs were regions of high
gas density and low shear.  Spurs are faint and difficult features to
detect.  Though we see some spur-like CO features, the correlation
between HII regions and spurs requires more sensitive CO observations.
A more practical approach would be to first trace the spurs with high
quality (resolution and sensitivity) optical-infrared color maps to
understand their locations, structure, and frequency.  Then a targeted
study with sensitive CO observations can help us better understand the
relationship between spurs, CO and \Ha.  Though possible, we feel that
it is unlikely that all star formation in bars occurs in spurs.
However we note that, if stars form in the spurs, then the radially
outwards offset can be naturally explained because the gas streamlines
diverge outwards on the trailing side of the dust lanes in this model.

In summary, both gas flow models can explain the CO-\Ha offsets but
neither is satisfactory, especially when we consider these and
previous observations of bars.  One possible improvement, especially
to the hydrodynamic models, could be the use of a multi-phased
molecular medium, consisting of a diffuse, gravitationally unbound
phase (e.g., \citealt{elm93,rand99,huette00}), and the usual dense,
bound phase made of giant molecular clouds (e.g.,
\citealt{scoville90,young91}).  In a bar, diffuse gas may be formed by
the disruption of clouds by the tidal field of a bar, or by off-center
cloud-cloud collisions \citep{huette00}.  At least in one bar, NGC
7479, tentative evidence of diffuse gas has been presented
\citep{huette00}.  The diffuse gas has a higher velocity width than
the bound phase, and so it may dominate the CO kinematics.  Perhaps
this is why studies of gas kinematics in the dust lane have been
consistent with the hydrodynamic models (e.g,
\citealt{regan97,regan99}).  And it may be that the star forming
component of the gas, the dense, bound clouds, behaves more like the
sticky particles in N-body simulations.

\subsection{Time Scales and Cloud Speeds in Context of Model Gas Flows}

Previous spiral arm studies found \Ha offset from the CO by a few
hundred pc \citep{vogel88,rand93,knapen96,loinard96}.  Vogel et
al. (1988) immediately recognized that this offset was too large given
that typical ``drift'' velocity of a cloud across a spiral arm is
$\sim$ 10 \kms.  They suggested that stars may be forming after a
``gestation'' period of a few million years following the compression
of clouds by the spiral density wave.

The situation is different in the bar environment.  In the
N-body/sticky particle model for bars, the orbit crowding is much more
severe than for a spiral arm shock, and it is non-trivial to calculate
an equivalent ``drift'' velocity across the bar dust lane.  The
offsets observed along most of the bar dust lanes (0--500 pc, \S
\ref{onedres}) are similar to those observed along spiral arms.
Therefore if the drift velocity in bars is similar to that in spiral
arms, similar ``gestation'' periods are necessary in the framework of
the N-body/sticky particle gas flow model.  However in one or two
locations along bars in NGC 2903 and NGC 3627, we measure offsets as
large as $\sim$ 800 pc.  Then for the star formation scenario
presented here in the framework of N-body/sticky particle gas flow
model, either unusually large gestation periods and/or large drift
velocities are required.  One observational test may be to measure
speeds of newly formed clusters using stellar absorption lines to
better understand the dynamics.  Another possibility is that for these
particular cases, our basic assumption that these stars formed in the
dust lanes is not applicable (see caveat in \S \ref{det}).

In the framework of the hydrodynamical models there is no ``drift''
velocity because all of the gas is redirected inwards by the shock.
Suppose the newly formed stars inherit the speed of their parent
molecular clouds.  Then to traverse the {\sl entire} range of offsets
seen in these bars (0--800 pc, \S \ref{onedres}), the velocity for a
stellar cluster must be 0--80 \kms, assuming a time scale of 10 Myr
for the \Ha peaks.  Individual HII regions have lifetimes $\sim$ 3 Myr
\citep{spitzer78} but since we do not resolve individual HII regions
at these distances, we assume a larger lifetime for the \Ha peaks
because they are most likely a collection of HII regions.

The relative speed of clouds entering the bar pattern is high. Though
difficult to quantify, a few direct measurements of bar pattern speeds
indicate values between 50--65 \kms\ kpc$^{-1}$
\citep{merrifield95,dehnen99, gerssen99}.  Hydrodynamic models which
try to reproduce observations use a bar pattern speed of 30--35 \kms\
kpc$^{-1}$ \citep{piner95}, but these bars are longer than typically
observed.  In any case, both the models and observations indicate that
bars end at $\sim$80\% of their co-rotation radius, consistent with
the findings of \citet{a92a,a92b}.

Using this empirical relationship, and deprojected bar lengths (Table
5 from \citealt{sheth02}), the bar pattern speeds in 5/6 galaxies were
measured using CO rotation curves by \citet{das01} to be: 68 \kms\
kpc$^{-1}$ (NGC 2903), 55 \kms\ kpc$^{-1}$ (NGC 3627), 35 \kms\
kpc$^{-1}$ (NGC 4321), 71 \kms\ kpc$^{-1}$ (NGC 5457), and 55 \kms\
kpc$^{-1}$ (NGC 6946).  These values are consistent with the direct
measurements of bar pattern speeds in other galaxies.  The pattern
speed of the bar in IC342 is difficult to estimate because the bar
parameters are not well-known.  We estimated a bar length of
$\sim$58$\arcsec$ (0.6 kpc).  For this bar length, the pattern speed
from the CO rotation curve is 196 \kms\ kpc$^{-1}$, an unrealistically
high value.  If instead we use a length of 4$\arcmin$ as suggested by
\citet{cros01}, and assume that the rotation curve remains flat at
$\sim$ 141 \kms, then the bar pattern speed would be a more reasonable
81 \kms\ kpc$^{-1}$.  We do not use IC 342 in the ensuing discussion.

The speed at which a molecular cloud enters is the difference between
the disk and bar pattern speeds.  Hence, at the mid-points of these
bars, cloud speed varies from 49--99 \kms.  So in the context of the
hydrodynamical models, if a stellar cluster forms from clouds upstream
of the shock, it would inherit a velocity of this magnitude.  The
velocity of the clouds upstream of the shock is primarily tangential
to the bar dust lane.  Hence, while the gas is redirected inwards down
the dust lane, newly formed stellar clusters continue on the original
orbits, and the observed range of offsets is consistent with the
expected cloud speeds.  However, we emphasize that the range of CO-\Ha
offsets in a bar is not correlated with the galacto-centric radius.
This suggests that star formation is not a simple function of the
parent cloud speed, or alternatively a function of the shock strength.
Consistent with this conclusion is the range in offsets observed in a
single bar which suggests that star formation depends on other
factors, e.g., gas surface density, and/or local shear.

\section{Conclusions}\label{conc}

We have investigated the distribution of molecular gas and star
forming regions in the bars of six spirals from the BIMA Survey of
Nearby Galaxies. Our main conclusions are as follows: \\

\noindent 1. The CO emission is brightest along the leading edges of
the bar. Weak spurs of CO emission are seen on the trailing side of
the dust lanes.  These spurs may play a role in star formation
upstream of the dust lane. At the bar ends, strong CO and \Ha emission
are seen on the trailing and leading sides; these may be the
beginnings of inner rings.

\noindent 2. The \Ha emission is distributed in compact and diffuse
structures. There are a few instances where the \Ha is coincident
with, or on the trailing side of the CO emission.  But the main result
is that in all six cases, the majority of the \Ha emission is offset
towards the leading side of the CO.

\noindent 3.  We quantify the offsets using a cross-correlation
analysis and find a range of 0--800 pc, with larger offsets in
stronger bars.  However, in a given bar there is a range of offsets
and there is no systematic pattern as a function of the
galacto-centric radius.

\noindent 4. In the two dimensional cross-correlation analysis, there
is a tendency for the \Ha emission to be offset radially outwards from
the CO emission.  However, this trend is less significant than the
azimuthal offsets because correlations in two-dimensions can be
between physically unrelated HII and CO regions.

\noindent 5. The observed CO-\Ha distributions may be explained by
either the N-body/sticky particle models or the hydrodynamic models
with different, plausible, star formation scenarios. In the context of
the N-body simulations, stars may form via cloud-cloud agglomeration
in the dust lanes. In the context of the hydrodynamic models, the
stars could form in dust spurs on the trailing side of the dust lane.
We suggest that addition of a two-phased or multi-phased molecular
medium can improve the agreement between these and previous
observations, and gas flow models in bars.

\centerline{\bf Acknowledgements}

This work would not have been possible without the rest of the SONG
team members (T. Wong, T. Helfer, L. Blitz and D. Bock) and the
dedicated observatory staff at Hat Creek and at the Laboratory for
Millimeter-wave Astronomy at the University of Maryland.  We thank
S. Aalto, S. H\"uttemeister, J. Kenney, E. Ostriker, N. Scoville,
E. Schinnerer, J. Stone, and T. Treu for invaluable and insightful
discussions about gas kinematics and star formation.  We are grateful
to N. Reddy and E. Schinnerer for their careful reading and helpful
comments which significantly improved this paper.

Research with the BIMA array is supported by NSF grant AST-9981289.
Support for the Laboratory for Millimeter-wave Astronomy is also
provided by the state of Maryland.  This research is also partially
funded by NSF grant AST-9981546.

\clearpage

\begin{figure}
\figcaption{NGC 2903: CO(1--0) emission contours from a BIMA SONG + 12m OTF
map are overlaid on top of continuum-subtracted \Ha emission. CO
contours are plotted at 2, 4, 6, 8, 10, 14, 18, 22, 30, 40, 50, and 75
$\times$ 2.2 Jy \kms. The \Ha image is not calibrated and is shown at
an arbitrary stretch. The dark ellipse shows the extent of the stellar
bar \citep{sheth01}. The curved arrow (bottom right) indicates the
direction of rotation, assuming that spiral arms trail.  \label{2903olay}}
\end{figure}
\begin{figure}
\figcaption{NGC 3627: Same as Figure \ref{2903olay}. Same CO contour
levels as in Figure \ref{2903olay} $\times$ 1.4 Jy \kms. The dashed
segments show an inner ring in CO and \Ha \citep{regan02}.
\label{3627olay}}
\end{figure}
\begin{figure}
\figcaption{NGC 4321: Same as Figure \ref{2903olay}. CO contour levels at
2, 3, 4, 6 ,8, 10, and 20 $\times$ 2.0 Jy \kms.\label{4321olay}}
\end{figure}
\begin{figure}
\figcaption{NGC 5457: Same as Figure \ref{2903olay}. CO contour levels at
1, 2, 3, 4, 6, 8, 10, and 20 $\times$ 1.8 Jy \kms. The CO and \Ha
emission in this galaxy have notably different distribution at the bar
ends, than in NGC 2903 or NGC 3627, in that the CO and \Ha emission
curves towards the leading side of the bar. This gas response 
is probably related to the properties of this late Hubble type bar. 
\label{5457olay}}
\end{figure}
\begin{figure}
\figcaption{NGC 6946: Same as Figure \ref{2903olay}. CO contour levels at 2,
4, 6, 8, 10, 15, 20, 25, 30, 35, 40, and 45 $\times$ 2.0 Jy
\kms. \label{6946olay}}
\end{figure}
\begin{figure}
\figcaption{IC 342: Same as Figure \ref{2903olay}. CO contour levels at 2,
3, 4, 6, 8, 10, 15, 20, 25, 30, 35, 40, 45 $\times$ 5.0 Jy
\kms. \label{ic342olay}}
\end{figure}
\begin{figure}
\figcaption{NGC 2903: The left column shows the deprojected CO (top)
and \Ha (bottom) images.  The solid circles indicate the boundaries
over which the cross-correlation was done.  The dashed circles are
spaced at 40,50, and 60$\arcsec$ from the nucleus.  The vertical panel
on the right hand side shows the maximum of the azimuthal
cross-correlation for 1$\arcsec$ radial bins.  The annotations refer
to the CO or HII regions labeled in Figure \ref{2903olay}.  The bottom
panel shows the two dimensional cross-correlation values.  The
contours, drawn at arbitrary levels, highlight peaks and valleys.
Notice that the cross-correlation values are highest for \Ha offset
azimuthally downstream (leading side) and radially outwards from the
CO.
\label{n2903all}}
\end{figure}
\begin{figure}
\figcaption{NGC 3627: Same as Figure \ref{n2903all}.  Diffuse
indicates correlation between diffuse \Ha and CO.  The notation
``n.C5'' refers to the northern side of the region C5.
\label{n3627all}}
\end{figure}
\begin{figure}
\figcaption{NGC 4321: Same as Figure \ref{n2903all}.  The arrows at 51
and 52$\arcsec$ indicate non-physical lags where the maximum of the
cross-correlation is dominated by a correlation between diffuse and
unrelated CO and \Ha emission in the annuli.
\label{n4321all}}
\end{figure}
\begin{figure}
\figcaption{NGC 5457: Same as previous figures. 
\label{n5457all}}
\end{figure}
\begin{figure}
\figcaption{NGC 6946: Same as previous figures.
\label{n6946all}}
\end{figure}
\begin{figure}
\figcaption{IC 342: Same as previous figures.
\label{ic342all}}
\end{figure}

\clearpage
\begin{deluxetable}{llrrrrrr}
\tablecaption{Properties of observed galaxies \label{tab1}}
\tablehead{\colhead{Galaxy}& \colhead{Type}& \colhead{RA\tablenotemark{a}}& \colhead{DEC}& \colhead{V$_{hel}$}& \colhead{PA}& \colhead{i}& \colhead{D} \\
\colhead{}& \colhead{RC3}& \colhead{J2000}& \colhead{J2000}& \colhead{\kms}& \colhead{deg}& \colhead{deg}& \colhead{Mpc}}
\startdata
IC 342 & SAB(rs)cd & 03:46:48.56 & +68:05:46.6 & 34 & 37 & 31.00 & 2.10\tablenotemark{1} \\ 
NGC 2903 & SAB(rs)bc & 09:32:10.05 & +21:30:02.0 & 556 & 17 & 61.40 & 7.30\tablenotemark{2} \\ 
NGC 3627 & SAB(s)b & 11:20:14.99 & +12:59:29.3 & 727 & 173 & 62.80 & 11.07\tablenotemark{3} \\ 
NGC 4321 & SAB(s)bc & 12:22:54.84 & +15:49:20.0 & 1571 & 30 & 31.7 & 16.1\tablenotemark{4} \\ 
NGC 5457 & SAB(rs)cd & 14:03:12.52 & +54:20:56.5 & 241 & 35 & 21.05 & 6.45\tablenotemark{5} \\ 
NGC 6946 & SAB(rs)cd & 20:34:52.33 & +60:09:14.2 & 48 & 35 & 31.66 & 6.40\tablenotemark{6} \\ 
\enddata
\tablenotetext{a}{Galaxy centers measured from either our optical data or from the 2MASS survey. Position angle (PA), inclination (i), Hubble type and the heliocentric velocity (V$_{hel}$) are from the RC3 catalog. References for the adopted distances (D) are given below.} \\
\tablenotetext{1}{\citet{karachen93}}
\tablenotetext{2}{\citet{planesas97}}
\tablenotetext{3}{\citet{saha99}}
\tablenotetext{4}{\citet{ferrarese96}}
\tablenotetext{5}{\citet{stetson98}}
\tablenotetext{6}{\citet{sharina97}}
\end{deluxetable}


\clearpage
\begin{thebibliography}{}

\bibitem[Athanassoula(1992a)]{a92a} Athanassoula, E. 1992, \mnras, 259, 328

\bibitem[Athanassoula(1992b)]{a92b} Athanassoula, E. 1992, \mnras, 259, 345

\bibitem[Benedict, Smith, \& Kenney(1996)]{benedict96} Benedict, 
F.~G., Smith, B.~J., \& Kenney, J.~D.~P.\ 1996, \aj, 111, 1861. 


\bibitem[Blitz(1993)]{blitz93} Blitz, L. 1993, Protostars and Planets
III, eds. E.H. Levy \& J.I. Lunine (Tucson:Univ. of Arizona Press), 12

\bibitem[Buta \& Combes(1996)]{buta96} Buta, R., \& Combes, F. 1996,
Fundamentals of Cosmic Physics, 17, 95

\bibitem[Buta et al.(2000)]{buta00a} Buta, R., Treuthardt, P.M., Byrd,
G.G., \& Crocker, D.A. 2000, \aj, 120, 1289

\bibitem[Buta \& Block(2000)]{buta00b} Buta, R., \& Block, D.L., 2000,
\apj, 550, 243

\bibitem[Combes \& Gerin(1985)]{combes85} Combes, F., \& Gerin,
M. 1985, \aap, 150, 327

\bibitem[Combes(1996)]{combes96} Combes, F. 1996, in {\em IAU
Colloquium 157, Barred Galaxies}, eds. R. Buta, D.A. Crocker, \&
B.G. Elmegreen, (San Francisco:ASP), 286

\bibitem[Crosthwaite et al.(2001)]{cros01} Crosthwaite, L.~P., 
Turner, J.~L., Hurt, R.~L., Levine, D.~A., Martin, R.~N., \& Ho, P.~T.~P.\ 
2001, \aj, 122, 797. 

\bibitem[Das et al.(2001)]{das01} Das, M., Teuben, P.~J., 
Vogel, S.~N., Harris, A., Regan, M.~W., Sheth, K., Helfer, T.~T., \& Thornley, M.~D.\ 2001, American Astronomical Society Meeting, 199, 5812. 

\bibitem[Dehnen(1999)]{dehnen99} Dehnen, W. 1999, \apjl, 524, 35

\bibitem[Downes, Reynaud, Solomon, \& Radford(1996)]{downes96} 
Downes, D., Reynaud, D., Solomon, P.~M., \& Radford, S.~J.~E.\ 1996, \apj, 
461, 186. 

\bibitem[Eckart et al.(1991)]{eckart91} Eckart, A., Cameron, M., 
Jackson, J.~M., Genzel, R., Harris, A.~I., Wild, W., \& Zinnecker, H.\ 
1991, \apj, 372, 67. 

\bibitem[Elmegreen(1993)]{elm93} Elmegreen, B.G. 1993, \apj 411, 170

\bibitem[Elmegreen \& Elmegreen(1985)]{elmelm85} Elmegreen, B.G., \&
Elmegreen, D.M. 1985, \apj, 288, 438

\bibitem[Eskridge et al.(2000)]{eskridge00} Eskridge, P., et
al. 2000, \aj, 119, 536

\bibitem[Ferrarese et al.(1996)]{ferrarese96} Ferrarese, L. et
al. 1996, \apjl, 468, 95

\bibitem[Friedli \& Benz(1993)]{friedli93}Friedli, D. \& Benz,
W. 1993, \aap, 268, 65

\bibitem[Garcia-Barreto et al.(1996)]{garcia96} Garcia-Barreto, J.A.,
\& Carrillo, R., Venegas, S. \& Escalante-Ramirez, B. 1996, in {\em
IAU Colloquium 157, Barred Galaxies}, eds. R. Buta, D.A. Crocker, \&
B.G. Elmegreen, (San Francisco:ASP), 76

\bibitem[Gerssen, Kuijken, \& Merrifield(1999)]{gerssen99} Gerssen,
J., Kuijken, K., \& Merrifield, M. R. 1999, \mnras, 306, 926

\bibitem[Gruendl(1996)]{gruendl96} Gruendl, R. 1996, Ph.D. Thesis,
University of Maryland

\bibitem[Handa et al.(1990)]{handa90} Handa, T., Nakai, N., 
Sofue, Y., Hayashi, M., \& Fujimoto, M.\ 1990, \pasj, 42, 1 

\bibitem[Helfer et al.(2002a)]{helfer02a} Helfer, T.T., Vogel, S.N.,
Lugten, J.B., \& Teuben, P.J. 2002, \pasp, 114, 350

\bibitem[Helfer et al.(2002b)]{helfer02b} Helfer, T.T., Thornley, M.D.,
Regan, M.W., Wong, T., Sheth, K., Vogel, S.N., Blitz, L., \& Bock,
D.C.-J. 2002, \apjs, submitted

\bibitem[Heller \& Shlosman(1994)]{heller94} Heller, C. H., \&
Shlosman, I. 1994, \apj, 424, 84

\bibitem[Ho, Filipenko, \& Sargent(1997)]{ho97} Ho, L.C., Filippenko,
A.V., \& Sargent, W.L.W. 1997, \apj, 487, 591

\bibitem[H\"uttemeister et al.(2000)]{huette00}H\"uttemeister, S.,
Aalto, S., Das M., \& Wall, W. F. 2000, A\&A, 363, 93

\bibitem[Karachentsev \& Tikhonov(1993)]{karachen93} Karachentsev,
I. D., \& Tikhonov, N. A. 1993, A \& AS, 100, 227

\bibitem[Kenney, Scoville, \& Wilson(1991)]{kenney91} Kenney, 
J.~D.~P., Scoville, N.~Z., \& Wilson, C.~D.\ 1991, \apj, 366, 432. 

\bibitem[Kenney, Carlstrom, \& Young(1993)]{kenney93} Kenney, 
J.~D.~P., Carlstrom, J.~E., \& Young, J.~S.\ 1993, \apj, 418, 687. 

\bibitem[Kennicutt(1998)]{kennicutt98} Kennicutt, R.C., 1998, \araa, 36

\bibitem[Kim \& Ostriker(2002)]{kim02} Kim, W.T., \& Ostriker, E. C.,
2002, \apj, 570, 132

\bibitem[Knapen \& Beckman(1996)]{knapen96} Knapen, J.H., \& Beckman, J.E., 1996, \mnras, 283, 251

\bibitem[Loinard et al.(1996)]{loinard96} Loinard, L., Dame, T. M., Koper, E., Lequeux, J., Thaddeus, P., \& Young, J. S. 1996, \apjl,  469,101

\bibitem[Lord \& Kenney(1991)]{lord91} Lord, S.D., \& Kenney, J.D.P.,
1991, \apjl, 381, 130

\bibitem[Martin\& Roy(1994)]{martin94} Martin, P., \& Roy, J. 1994,
\apj, 424, 599

\bibitem[Martin \& Friedli(1997)]{martin97} Martin, P., \& Friedli,
D. 1997, \aap, 326, 449

\bibitem[Martinet \& Friedli(1997)]{martinet97} Martinet, L., \&
Friedli, D. 1997, \aap, 323, 363

\bibitem[Merrifield \& Kuijken(1995)]{merrifield95} Merrifield, M.R.,
\& Kuijken, K. 1995, \mnras, 274, 933

\bibitem[Miyashiro(1982)]{miyashiro82} Miyashiro, G. M. 1982, Zodiac
User's Manual, 2nd ed.

\bibitem[Norman, Sellwood \& Hasan(1996)]{norman96} Norman, C. A.,
Sellwood, J. A., \& Hasan, H. 1996, \apj, 462, 114

\bibitem[Ohta et al.(1986)]{ohta86} Ohta, K., Sasaki, M., Saito,
M. 1986, PASJ, 38, 677

\bibitem[Ondrechen(1985)]{ondrechen85} Ondrechen, M.~P.\ 1985, \aj, 
90, 1474

\bibitem[Petitpas \& Wilson(1998)]{1998ApJ...503..219P} Petitpas, G.~R.~\& Wilson, C.~D.\ 1998, \apj, 503, 219. 

\bibitem[Phillips(1996)]{phillips96} Phillips, A.C. 1996, in IAU
Colloquium 157, Barred Galaxies, eds. R. Buta, D.A. Crocker, \&
B.G. Elmegreen, (San Francisco:ASP), 44

\bibitem[Piner, Stone \& Teuben(1995)]{piner95} Piner, B.G., Stone,
J.M., \& Teuben, P. J. 1995, \apj, 449, 508

\bibitem[Planesas et al.(1997)]{planesas97} Planesas, P., Colina,
L. \& Perez-Olea,D. 1997, \aap, 325, 81

\bibitem[Rand(1993)]{rand93} Rand, R.J., 1993, \apj, 404, 593

\bibitem[Rand(1995)]{rand95} Rand, R.J., 1995, \aj, 109, 2444

\bibitem[Rand, Lord \& Higdon(1999)]{rand99} Rand, R.~J., 
Lord, S.~D., \& Higdon, J.~L.\ 1999, \apj, 513, 720 

\bibitem[Regan \& Vogel(1995)]{regan95} Regan, M.W., \& Vogel,
S.N. 1995, \apjl, 452, 21

\bibitem[Regan, Vogel \& Teuben(1997)]{regan97} Regan, M. W., Vogel,
S.N., Teuben, P.J. 1997, \apjl, 482, 135

\bibitem[Regan \& Elmegreen(1997)]{reganelm97} Regan, M.W., \&
Elmegreen, D.M. 1997, \aj, 114, 965

\bibitem[Regan, Sheth \& Vogel(1999)]{regan99} Regan, M. W., Sheth,
K., \& Vogel, S.N. 1999, \apj, 526, 97

\bibitem[Regan et al.(2001)]{regan01} Regan, M.W., Helfer, T.T.,
Thornley, M.D., Sheth, K., Wong, T., Vogel, S.N., Blitz, L., \& Bock,
D.C.-J. 2001, \apj, 561, 218

\bibitem[Regan et al.(2002)]{regan02} Regan, M.W., Sheth, K., Vogel,
S.N., \& Teuben, P.J. 2002, \apj, in press

\bibitem[Roberts, Huntley \& van Albada(1979)]{roberts79} Roberts,
W. W., Jr., Huntley, J. M., \& van Albada, G. D., 1979, \apj, 233, 67

\bibitem[Roy \& Belley(1993)]{roy93} Roy, J.~\& Belley, J.\ 
1993, \apj, 406, 60. 

\bibitem[Saha et al.(1999)]{saha99} Saha, A., Sandage, A., Tammann,
G. A., Labhardt, L., Macchetto, F. D., Panagia, N., 1999, \apj, 522,
802

\bibitem[Sakamoto et al.(1995)]{sakamoto95} Sakamoto, K., Okumura, 
S., Minezaki, T., Kobayashi, Y., \& Wada, K.\ 1995, \aj, 110, 2075. 

\bibitem[Sakamoto et al.(1999)]{sakamoto99} Sakamoto, K., Okumura,
S. K., Ishizuki, S., Scoville, N. Z. 1999, \apj, 525, 691

\bibitem[Sault, Teuben \& Wright(1995)]{sault95} Sault, R. J.,
Teuben, P. J., \& Wright, M. C. H. 1995, {\em Astronomical Data
Analysis Software and Systems IV, ASP Conference Series, Volume \#77},
eds. R.A. Shaw, H.E. Payne, and J.J.E. Hayes, 4, 433

\bibitem[Sempere \& Garcia-Burillo(1997)]{sempere97} Sempere, M. J., \& 
Garcia-Burillo, S., 1997, \aap, 325, 769

\bibitem[Scoville(1990)]{scoville90} Scoville, N.Z. 1990, in the {\em Proceedings of The Evolution of the Interstellar Medium}, ASP: San Francisco, CA (A91-55426 24-90), p. 49-61.

\bibitem[Sellwood \& Wilkinson (1993)]{sellwood93} Sellwood, J. A., \&
Wilkinson, A., 1993, Rep. Prog. Phys, 56, 173

\bibitem[Sheth et al.(1998)]{sheth98} Sheth, K.~et al.\ 1998, 
American Astronomical Society Meeting, 193, 7012. 


\bibitem[Sheth et al.(2000)]{sheth00} Sheth, K., Regan, M.W., Vogel,
S.N., \& Teuben, P.J. 2000 \apj, 532, 221

\bibitem[Sheth(2001)]{sheth01} Sheth, K. 2001, Ph.D. Thesis,
University of Maryland, \\
http://www.astro.caltech.edu/$\sim$kartik/thesis.html

\bibitem[Sheth et al.(2002)]{sheth02} Sheth, K., et al., 2002, \apj,
in preparation.

\bibitem[Spitzer(1978)]{spitzer78} Spitzer, Jr. L., 1978, in {\em Physical
Processes in the Interstellar Medium}, (New York: Wiley)

\bibitem[Stetson et al.(1998)]{stetson98} Stetson, P.B. et al., 1998,
\apj, 508, 491

\bibitem[Sharina et al.(1997)]{sharina97} Sharina, M. E.,
Karachentsev, I. D., \& Tikhonov, N. A. 1997, Astronomy Letters,
23,373

\bibitem[Shopbell(1997)]{shopbell97} Shopbell, P. L. 1997, Zodiac+
User's Manual, 1st ed.

\bibitem[Vogel et al.(1995)]{vogel95}Vogel, S.N., Weymann, R., Rauch,
M., \& Hamilton, T. 1995, \apj, 441, 162

\bibitem[Vogel, Kulkarni \& Scoville(1988)]{vogel88} Vogel, S.N.,
Kulkarni, S.R., \& Scoville, N.Z. 1988, \nat, 402

\bibitem[Young \& Scoville(1991)]{young91} Young, J.S. \& Scoville, N.Z. 1991 \araa, 29, 581
\end{thebibliography}
\end{document}